\documentclass[twocolumn,floatfix,superscriptaddress,a4paper,showpacs,showkeys,nofootinbib,reprint,prc]{revtex4-1}
\usepackage{epsfig}
\usepackage{latexsym}
\usepackage[colorlinks=true,linktocpage=true,linkcolor=blue,citecolor=blue,allcolors=blue]{hyperref}
\usepackage{url}
\usepackage[utf8]{inputenc}
\usepackage{enumerate}
\usepackage{color}
\usepackage{xcolor}
\usepackage{amsmath,amssymb}

\usepackage[english]{babel}
\usepackage{hyperref}
\usepackage{url}
\usepackage{needspace}
\hypersetup{
   colorlinks=true, 
  linktoc=all,     
  linkcolor=blue,  
}

\begin{document}

\title{Possible explanation of the irregular energy dependence of the rapidity width of $\phi$ mesons observed in Pb+Pb collisions}

\author{Tom Reichert}
\affiliation{Institut f\"{u}r Theoretische Physik, Goethe-Universit\"{a}t Frankfurt, Max-von-Laue-Str. 1, D-60438 Frankfurt am Main, Germany}
\affiliation{Frankfurt Institute for Advanced Studies (FIAS), Ruth-Moufang-Str. 1, D-60438 Frankfurt am Main, Germany}
\affiliation{Helmholtz Research Academy Hesse for FAIR (HFHF), GSI Helmholtzzentrum f\"ur Schwerionenforschung GmbH, Campus Frankfurt, Max-von-Laue-Str. 12, 60438 Frankfurt am Main, Germany}

\author{Jan Steinheimer}
\affiliation{Frankfurt Institute for Advanced Studies (FIAS), Ruth-Moufang-Str. 1, D-60438 Frankfurt am Main, Germany}

\author{Marcus Bleicher}
\affiliation{Institut f\"{u}r Theoretische Physik, Goethe-Universit\"{a}t Frankfurt, Max-von-Laue-Str. 1, D-60438 Frankfurt am Main, Germany}
\affiliation{GSI Helmholtzzentrum f\"ur Schwerionenforschung GmbH, Planckstr. 1, D-64291 Darmstadt, Germany}
\affiliation{Helmholtz Research Academy Hesse for FAIR (HFHF), GSI Helmholtzzentrum f\"ur Schwerionenforschung GmbH, Campus Frankfurt, Max-von-Laue-Str. 12, 60438 Frankfurt am Main, Germany}

\date{\today}

\begin{abstract}
Experimental data from the NA49 collaboration show an unexpectedly steep rise of the rapidity width of the $\phi$ meson as function of beam energy, which was suggested as possible interesting signal for novel physics. In this work we show that the Ultra-relativistic Quantum-Molecular-Dynamics (UrQMD) model is able to reproduce the shapes of the rapidity distributions of most measured hadrons and predicts a common linear increase of the width for all hadrons.
Only when following the exact same analysis technique and experimental acceptance of the NA49 and NA61/SHINE collaborations, we find that the extracted value of the rapidity width of the $\phi$ increases drastically for the highest beam energy. We conclude that the observed steep increase of the $\phi$ rapidity width is a problem of limited detector acceptance and the simplified Gaussian fit approximation.
\end{abstract}

\maketitle

\section{Introduction}
A promising observable for the understanding of strongly interacting matter created in heavy ion collisions is the production of $\phi$ mesons. Being a $\Bar{s}s$ state with zero net-strangeness it has a small hadronic cross section and may therefore transport information directly from hadronisation. Being at the same time a multi-strange hadron and net strangeness neutral it is further sensitive to different realizations of strangeness suppression in peripheral collisions, either via canonical strangeness suppression or via a $\gamma_s$ strangeness suppression factor. It has been further found that many transport simulations have substantial problems to describe the yields of $\phi$ mesons, which may hint to hitherto unknown production channels. For the production of $\phi$ mesons in hadronic and partonic matter many mechanisms have been proposed, such as thermal production \cite{Vovchenko:2015idt}, OZI suppressed reactions \cite{Chung:1997mp}, kaon coalescence \cite{Ko:1991kw}, catalytic reactions \cite{Kolomeitsev:2009yn,Tomasik:2011aa} or from heavy resonance decays \cite{Steinheimer:2015sha}.

The production of $\phi$ mesons has been studied extensively in experiments at the CERN-LHC \cite{ALICE:2014jbq}, RHIC \cite{PHENIX:2004spo,STAR:2008inc,STAR:2008bgi}, CERN-SPS \cite{NA49:2000jee,NA49:2008goy,NA61SHINE:2019gqe}, BNL-AGS \cite{E917:2003gec} and at GSI \cite{FOPI:2002csf,HADES:2009lnd,HADES:2017jgz}. Especially deep sub-threshold production of strangeness in the $\phi$ and $\Xi$ hadrons has been subject of low energy and large baryon density experiments e.g. at HADES \cite{HADES:2009lnd,HADES:2009mtu,HADES:2017jgz} accompanied by theoretical efforts \cite{Tomasik:2011aa,Steinheimer:2015sha}.

\begin{figure} [t]
    \centering
    \includegraphics[width=\columnwidth]{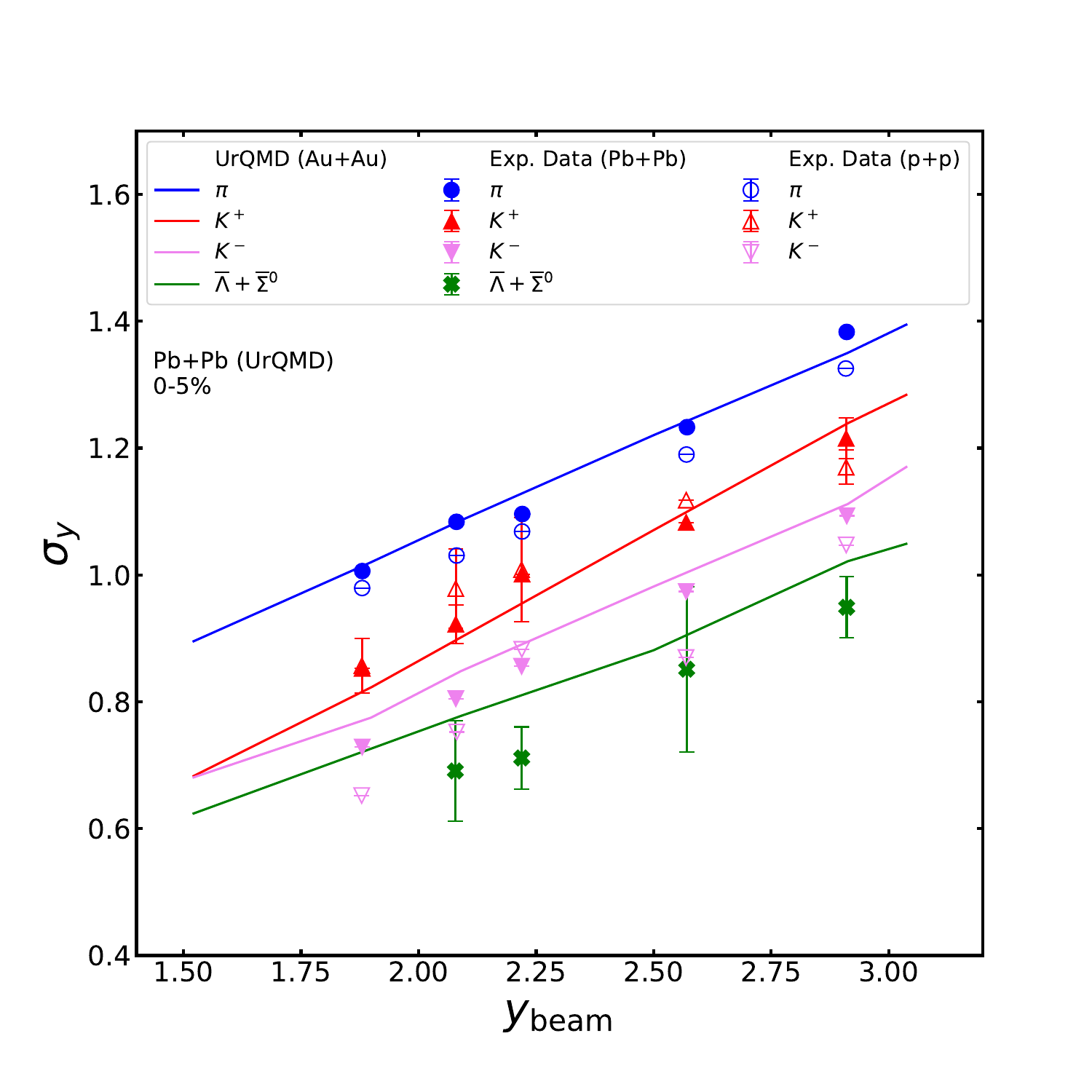}
    \caption{[Color online] Comparison between experimental results \cite{NA49:2000jee,NA61SHINE:2019gqe,NA61SHINE:2013tiv,Pulawski:thesis,NA49:2008goy,NA49:2002pzu,NA49:2007stj,NA49:2003bok} of the rapidity widths of $\pi$ (blue), $K^+$ (red), $K^-$ (magenta) and $\overline{\Lambda}+\overline{\Sigma}^0$ (green) in p+p (open symbols) and Pb+Pb (filled symbols) collisions and UrQMD simulations for Pb+Pb reactions (solid lines) in the energy range from $E_{\mathrm{lab}}=20$AGeV to $E_{\mathrm{lab}}=158$AGeV.}
    \label{fig:sigma_y_motivation}
\end{figure}

The NA49 collaboration at the CERN-SPS has reported on the rapidity widths of several hadrons ranging from $\pi$, $K^\pm$ to $\Bar{\Lambda}$ and to the $\phi$ meson for central Pb+Pb collisions in the energy range from $E_{\mathrm{lab}}=20$AGeV to $E_{\mathrm{lab}}=158$AGeV \cite{NA49:2008goy}. Recently, additional measurements in p+p collisions in the same energy range have been reported by the NA61/SHINE experiment \cite{NA61SHINE:2019gqe}. The general observations for $\pi$, $K^\pm$ and $\Bar{\Lambda}$ was that the rapidity widths increase linearly as a function of the beam rapidity both in Pb+Pb and in p+p collisions in the investigated energy regime and that the rapidity width in Pb+Pb and p+p are very similar to each other. These results are summarized as open (p+p) and filled (Pb+Pb) symbols in figure \ref{fig:sigma_y_motivation}. The $\phi$ seems to follow this linear trend in p+p collisions, but shows a stark difference in Pb+Pb collisions. For Pb+Pb collisions the data suggest a dramatic deviation from the trend observed for all other hadrons, namely a much broader rapidity width of the $\phi$ meson for the highest collision energy.

In this article we aim to explore the reason for this apparent broadening of the rapidity width of the $\phi$ in Pb+Pb collisions at 158A GeV using microscopic transport simulations. We will first calculate the rapidity distributions and widths directly from the UrQMD simulations and then proceed with the analysis following the experimental procedure to extract $\sigma_y$ of the $\phi$ meson.

\section{The UrQMD model and resonance reconstruction}
The Ultra-relativistic Quantum Molecular Dynamics model (UrQMD) \cite{Bass:1998ca,Bleicher:1999xi,Bleicher:2022kcu} is based on the covariant propagation of hadrons and their resonances in phase space. As a QMD-type model it keeps track of all n-body correlations providing a realistic time evolution for the n-body phase space distribution function. In its current version UrQMD includes mesonic and baryonic hadrons and resonances up to 4 GeV in mass. Their binary interactions are modeled via measured or derived cross sections, while additional QMD potentials can be switched on but will not be used in the present work.

Experimentally, unstable resonances like the $\phi$ are reconstructed from the invariant masses of their decay products. This means, from the invariant mass distributions of the measured $\Bar K K$-pairs, the yield of the $\phi$ is obtained by fitting a Breit-Wigner distribution in the mass range of the $m_\phi$ after subtracting the uncorrelated background. Thus, only $\phi$ mesons where both decay daughter particles escape without interaction from the fireball can be reconstructed experimentally. Therefore, in the simulation, measurable resonances are identified by tracing their daughter particles and counting only those $\phi$ mesons for which both daughter particles escape without further interactions. This method is well established and has been extensively tested at various collision energies and resonances \cite{Bleicher:2002dm, Bleicher:2002rx,Vogel:2005pd,Knospe:2015nva,Reichert:2019lny,Knospe:2021jgt,Reichert:2022uha}.

\section{Results}
To establish the baseline for the investigation, we first compare the rapidity widths of  $\pi$, $K^\pm$ and $\Bar{\Lambda}$ in Pb+Pb reactions with those presented in the NA61/SHINE article \cite{NA61SHINE:2019gqe}. For this purpose we calculate the rapidity distributions of $\pi$, $K^+$, $K^-$ and $\overline{\Lambda}+\overline{\Sigma}^0$ and extract the width of the rapidity distribution $\sigma_y$ from the simulated data precisely by $\sigma_y^2=\langle y^2\rangle - \langle y\rangle^2$. Fig. \ref{fig:sigma_y_motivation} shows the comparison of the rapidity widths calculated from UrQMD (solid lines) to the experimental data \cite{NA49:2000jee,NA61SHINE:2019gqe,NA61SHINE:2013tiv,Pulawski:thesis,NA49:2008goy,NA49:2002pzu,NA49:2007stj,NA49:2003bok} of $\pi$ (blue), $K^+$ (red), $K^-$ (magenta) and $\overline{\Lambda}+\overline{\Sigma}^0$ (green) in Pb+Pb (filled symbols) collisions. The experimental data in p+p collisions are also shown as open symbols.
\begin{figure} [t]
    \centering
    \includegraphics[width=\columnwidth]{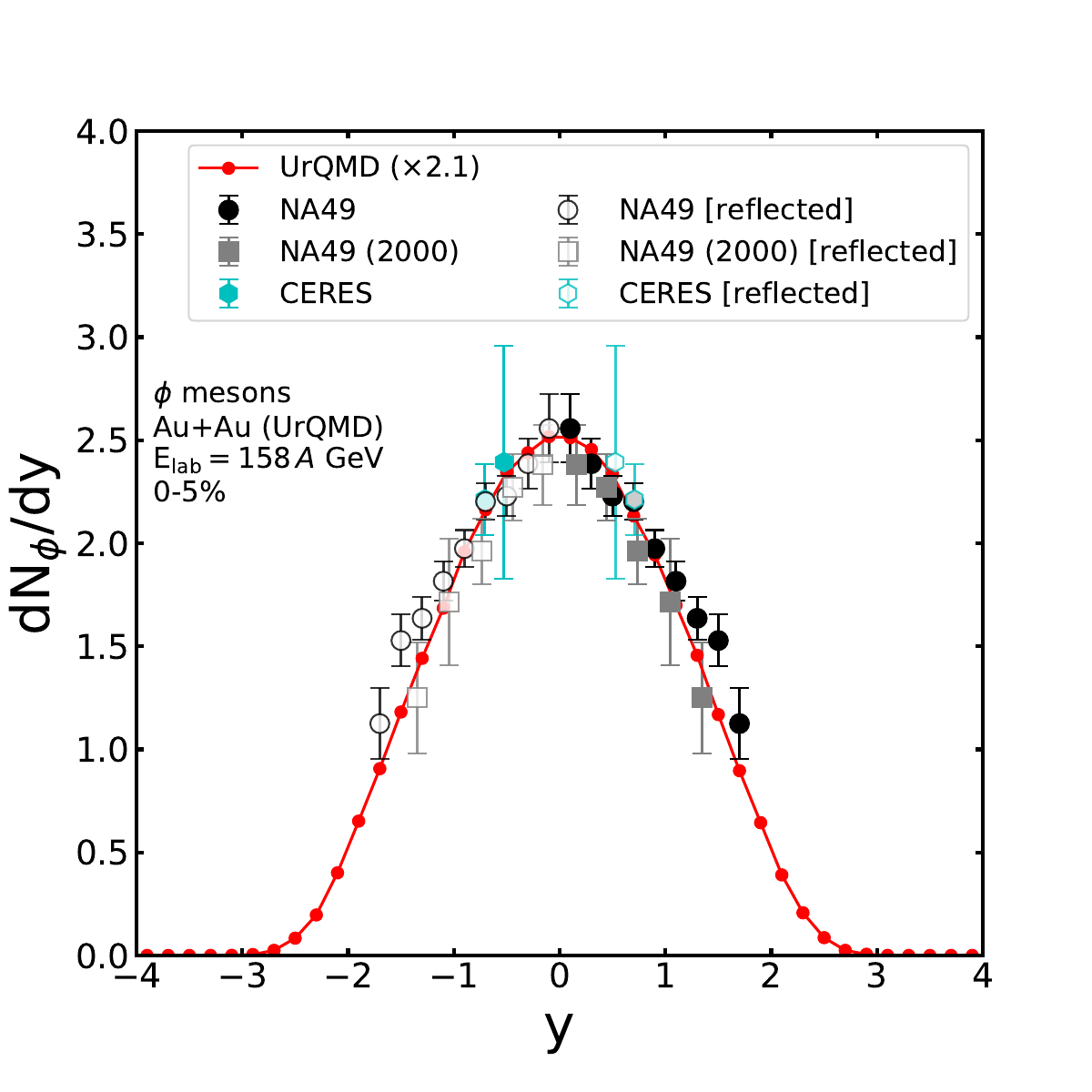}
    \caption{[Color online] Rapidity distributions of reconstructable $\phi$ mesons in 0-5\% central Pb+Pb collisions at E$_\mathrm{lab}=158 A$ GeV (red, scaled by 2.1) from UrQMD in cascade mode. The symbols represent experimental data taken at $E_\mathrm{lab}=158A$ GeV ($\sqrt{s_\mathrm{NN}}=17.3$ GeV) by NA49 \cite{NA49:2008goy} (black circles), earlier by NA49 \cite{NA49:2000jee} (grey squares) and by CERES \cite{CERES:2005wwe} (blue hexagons). The reflected experimental data is shown by brighter symbols and colors.}
    \label{fig:dNdy}
\end{figure}
One observes that the transport model calculations of $\sigma_y$ agree very well with the experimentally measured data for all investigated hadron species. Both model and data show a linear increase of the width of the rapidity distributions with increasing beam rapidity, with a roughly constant slope across all hadron species. 

As reported in \cite{NA49:2008goy} the $\sigma_y$ of the $\phi$ seems to not follow this behavior, especially close to $\sqrt{s_\mathrm{NN}}\approx20$ GeV.

Before calculating the width of the rapidity distribution of the $\phi$, the rapidity density should be compared directly between the model calculation and the experimental data to confirm the ability of the model to describe the shape of the rapidity distribution. Figure \ref{fig:dNdy} shows the rapidity distribution of reconstructable $\phi$ mesons in 0-5\% central Pb+Pb collisions at 158 AGeV (red, scaled by a factor 2.1) from UrQMD. The symbols represent experimental data by the NA49 collaboration \cite{NA49:2008goy} (black circles), earlier data by the NA49 \cite{NA49:2000jee} (grey squares) and by the CERES experiment \cite{CERES:2005wwe} (blue hexagons). The reflected experimental data is shown by lighter/open symbols.

We note that the UrQMD simulations have been scaled for better comparison by a factor of 2.1 (obtained through a fit to the available data) such that the integrated yield of the $\phi$ matches the experimental value. This scaling does, however, not affect the shape and width of the calculated rapidity distribution.

The shape of the rapidity density of $\phi$ mesons calculated with UrQMD describes the measured rapidity density very well, especially in terms of the width. This is also quantified in the reduced chi-squared $\tilde{\chi}^2 \equiv \chi^2/d.o.f.$ of the UrQMD-Data comparison which evaluates to $\tilde{\chi}_\mathrm{UrQMD}^2 = 0.16$ for UrQMD and to $\tilde{\chi}_\mathrm{Gauss}^2 = 0.10$ for a single Gaussian fit to the data. The quality of both fits is hence very good and both can be considered a good description of the available data.\footnote{Although the overall normalization of the UrQMD curve underestimates the measured $\phi$ yields, the $\chi^2/d.o.f.$ of the scaled curve suggests that the kinematics of the $\phi$ production are handled properly, hence allowing to reliably calculate the width of the rapidity distribution.}
\begin{figure} [t]
    \centering
    \includegraphics[width=\columnwidth]{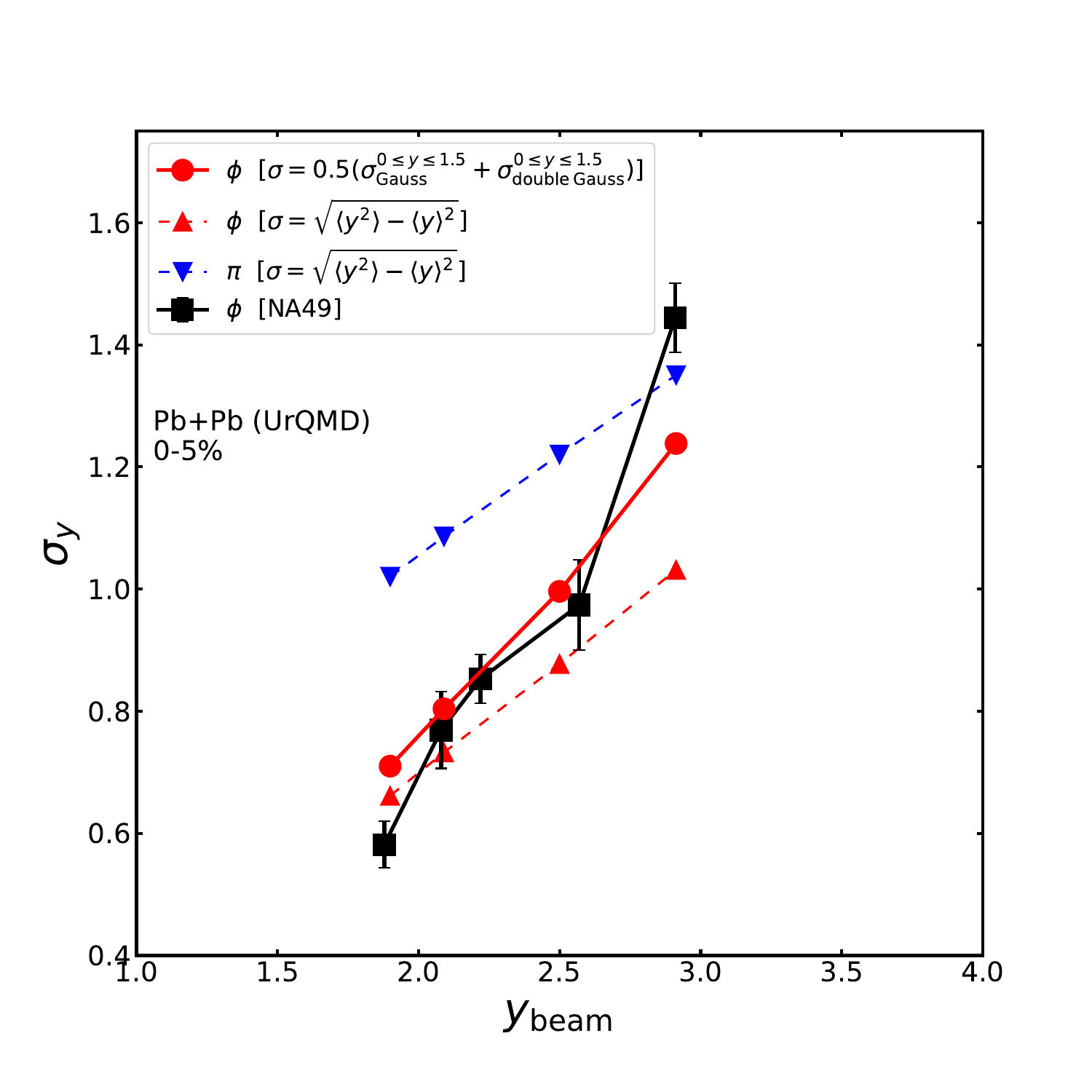}
    \caption{[Color online] The rapidity width $\sigma_y$ as a function of the beam rapidity of the $\phi$ calculated using the experimental technique (red filled circles) and calculated directly from the data set (red triangles-up) in 0-5\% central Pb+Pb collisions at different energies from UrQMD. The UrQMD calculations of the width of the $\pi$ (blue triangles-down) are shown as a comparison. Experimental results for the $\phi$ \cite{NA49:2000jee,NA49:2008goy} in Pb+Pb are shown as black squares with error bars.}
    \label{fig:sigma_y}
\end{figure}
To calculate the width of the $\phi$ rapidity density $\sigma_y^\phi$ two different methods can be employed: 
\begin{enumerate}
    \item The width is directly calculated, assuming full rapidity acceptance, as $\sigma_y = \sqrt{\langle (\delta y)^2 \rangle} = \sqrt{\langle y^2 \rangle - \langle y \rangle^2}$ from the simulated data set as it was done for the comparison of the $\pi$, $K^\pm$ and $\overline{\Lambda}$ widths in figure \ref{fig:sigma_y_motivation}.
    \item We amploy the method used by the NA49 and NA61/SHINE collaborations, i.e. we fit the rapidity density with a Gaussian \eqref{eq:gauss} and a double Gaussian\footnote{Note, that in case of a double Gaussian the width of the distribution is not equal to the parameter in the denominator, hence we renamed it as $\varpi$ to avoid confusion.} \eqref{eq:doublegauss} in the acceptance of the NA49 and NA61/SHINE detectors, i.e. $0\leq y\leq1.5$. Following the experimental analysis, we then extract a width from each fit and take the average width from both fits as the result, i.e. $\sigma_\mathrm{Exp.} = 0.5\,(\sigma_\mathrm{Fit\,Gauss}^{0\leq y\leq1.5}+\sigma_\mathrm{Fit\,double\,Gauss}^{0\leq y\leq1.5})$
\end{enumerate}
\begin{figure} [t]
    \centering
    \includegraphics[width=\columnwidth]{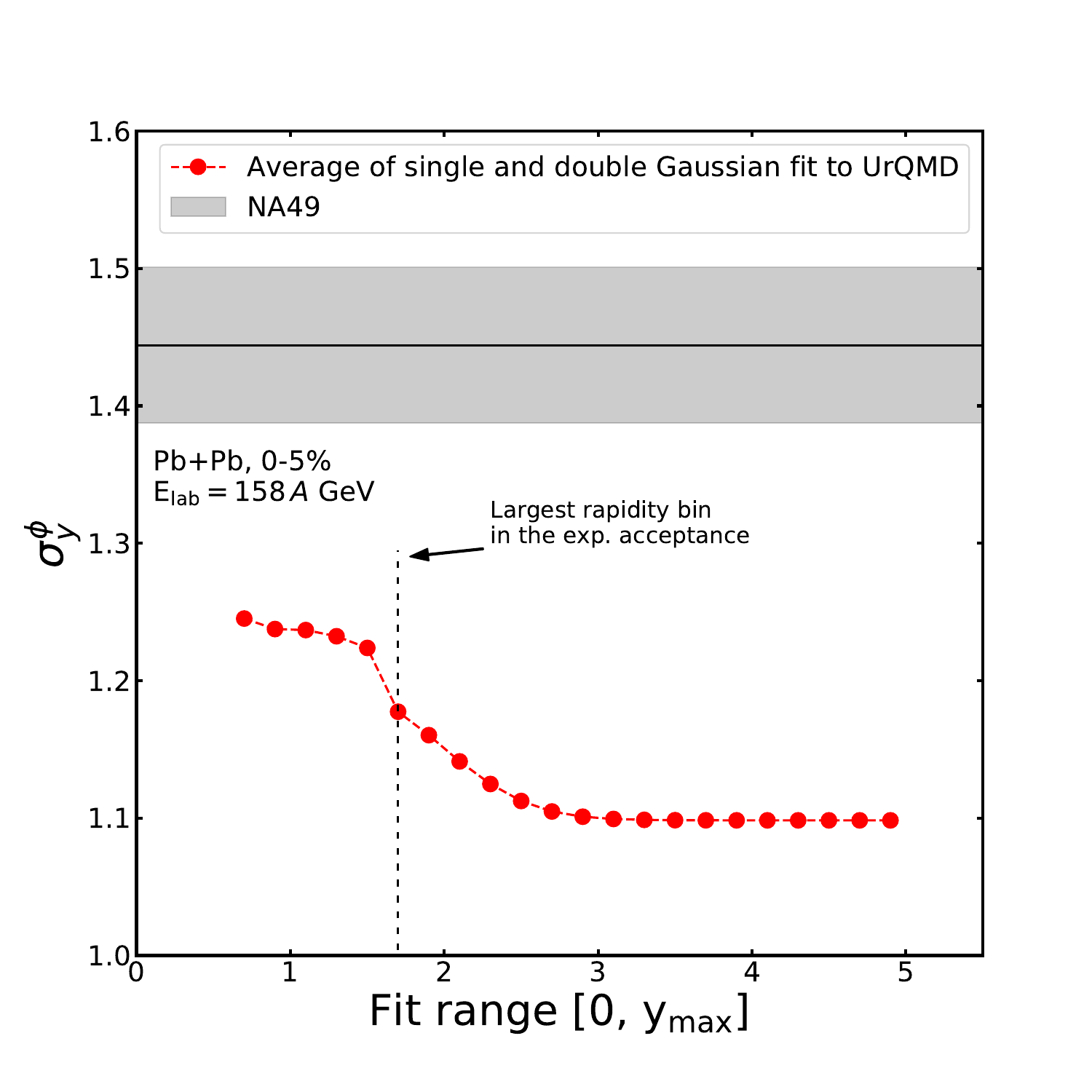}
    \caption{[Color online] The rapidity width $\sigma_y$ of $\phi$ meson calculated as the average width of the fit of a single Gaussian and a double Gaussian in the rapidity range $0\leq y \leq y_\mathrm{max}$ to the $\phi$ rapidity density in 0-5\% central Pb+Pb collisions at E$_\mathrm{lab}=158 A$ GeV from UrQMD. The experimental results \cite{NA49:2000jee,NA49:2008goy} in Pb+Pb are shown as a black line with error band.}
    \label{fig:sigma_y_range}
\end{figure}
\begin{align}\label{eq:gauss}
    \frac{\mathrm{d}N}{\mathrm{d}y} &\propto \exp\left[ \frac{y^2}{2\sigma^2} \right]
\end{align}
\begin{align}\label{eq:doublegauss}
    \frac{\mathrm{d}N}{\mathrm{d}y} &\propto \exp\left[ \frac{(y+a)^2}{2\varpi^2} \right] + \exp\left[ \frac{(y-a)^2}{2\varpi^2} \right]
\end{align}

Fig. \ref{fig:sigma_y} compares the calculated rapidity widths $\sigma_y$ (red triangles-up) and $\sigma_\mathrm{Exp.}$ (red filled circles) of the $\phi$ from UrQMD, in 0-5\% central Pb+Pb collisions at different energies, as a function of the beam rapidity. The UrQMD calculations of the width of the $\pi$ (blue triangles-down) are shown as a comparison. Experimental results for the $\phi$ \cite{NA49:2000jee,NA49:2008goy} in Pb+Pb are shown as black squares with error bars.

Though the experimental value of the width $\sigma_y$ of the $\phi$ is still larger than the widths of the $\phi$, calculated from the simulated UrQMD events, one notices that the width of the rapidity density of the $\phi$ calculated by following exactly the experimental technique shows also a much larger value than the direct calculation of $\sigma_y$. One should also keep in mind that this large variation of the width occurs for three different, but statistically very likely (i.e. all have low $\chi^2$), descriptions of the data.
The UrQMD model calculation of $\sigma_\mathrm{Exp.}$ of the $\phi$ meson employing the experimental technique comes very close to the experimental value though the true value of the width $\sigma_{y}$ shows the same linear trend with the same slope as the other hadrons. 


To better understand the systematics of the strong increase of $\sigma_\mathrm{Exp.}$, Fig. \ref{fig:sigma_y_range} shows the rapidity width $\sigma_\mathrm{Exp.}$ of $\phi$ mesons as a function of the reduced rapidity range $0\leq y \leq y_\mathrm{max}$, which is used as input for the Gaussian fit, from UrQMD as red circles. The experimental results \cite{NA49:2000jee,NA49:2008goy} in Pb+Pb are shown as a black line with error band.

The rapidity width $\sigma_\mathrm{Exp.}$ from the UrQMD calculation shows a strong dependence on the considered fit range (i.e. the acceptance of the experimental detector). Smaller acceptance windows artificially increase the extracted rapidity width. This can be attributed to the extrapolation of the employed fit function to large forward and backward rapidities, which deviates from the actual tails of the simulated distributions, simply because the distribution does not exactly resemble a double-Gaussian. This phenomenon is not observed at lower collision energies because the acceptance in beam rapidity allowed the single or double Gaussian fit to capture the whole distribution, constraining it also at higher forward and backward rapidities.

Again, we want to emphasize that all fits will provide a low $\Tilde{\chi}^2$ as the fit usually does not put much emphasis on the tails of the distributions. In conclusion it is not clear whether the systematic error, given by the experiment, is sufficient to capture the uncertainties arising from fitting a double-Gaussian in an incomplete acceptance. Therefore also the interpretation of the steep increase of the width as a signal for new physics is questionable.

\section{Conclusion}
In the present article we have employed the Ultra-relativistic Quantum Molecular Dynamics (UrQMD) model to calculate 0-5\% central Pb+Pb collisions and extract the rapidity distributions of reconstructable $\phi$ mesons. The NA49 and NA61 collaborations observed a strong increase of the width of the rapidity density of $\phi$ mesons around $\sqrt{s_\mathrm{NN}} \approx 20$ GeV. We have demonstrated that this increase in $\sigma_y$ may arise as an artifact from the limited rapidity coverage of the detector being fitted with a function which may not be able to capture the tails of the distributions correctly. The rapidity width calculated directly from the simulated data set of the $\phi$ follows the same linear trend as seen for the $\pi$, $K^\pm$ and $\overline{\Lambda}+\overline{\Sigma}^0$ while the width extracted within the experimental acceptance and employing the same fit function shows a very different systematic. The interpretation of 'novel' physics is disfavored by this investigation and caution is advised.

\begin{acknowledgments}
T.R. acknowledges support through the Main-Campus-Doctus fellowship provided by the Stiftung Polytechnische Gesellschaft Frankfurt am Main (SPTG). 
The authors acknowledge for the support the European Union's Horizon 2020 research and innovation program under grant agreement No 824093 (STRONG-2020). 
The computational resources for this project were provided by the Center for Scientific Computing of the GU Frankfurt and the Goethe-HLR.
\end{acknowledgments}



\end{document}